\documentclass[aps,prd,twocolumn,groupedaddress,showpacs,preprintnumbers,draft]
{revtex4}
\usepackage{epsf} 
\begin{document}
\preprint{MCGILL-22-05}
\def\Box{\nabla^2}  
\def\ie{{\em i.e.\/}}  
\def\eg{{\em e.g.\/}}  
\def\etc{{\em etc.\/}}  
\def\etal{{\em et al.\/}}  
\def\S{{\mathcal S}}  
\def\I{{\mathcal I}}  
\def\mL{{\mathcal L}}  
\def\H{{\mathcal H}}  
\def\M{{\mathcal M}}  
\def\N{{\mathcal N}} 
\def\O{{\mathcal O}} 
\def\cP{{\includegraphics[]
\mathcal P}} 
\def\R{{\mathcal R}}  
\def\K{{\mathcal K}}  
\def\W{{\mathcal W}} 
\def\mM{{\mathcal M}} 
\def\mJ{{\mathcal J}} 
\def\mP{{\mathbf P}} 
\def\mT{{\mathbf T}} 
\def\mR{{\mathbf R}}
\def\mS{{\mathbf S}}
\def\mX{{\mathbf X}}
\def\mZ{{\mathbf Z}}
\def\eff{{\mathrm{eff}}}  
\def\Newton{{\mathrm{Newton}}}  
\def\bulk{{\mathrm{bulk}}}  
\def\brane{{\mathrm{brane}}}  
\def\matter{{\mathrm{matter}}}  
\def\tr{{\mathrm{tr}}}  
\def\nr{{\mathrm{normal}}}  
\def\implies{\Rightarrow}  
\def\half{{1\over2}}  
\newcommand{\da}{\dot{a}}
\newcommand{\db}{\dot{b}}
\newcommand{\dn}{\dot{n}}
\newcommand{\dda}{\ddot{a}}
\newcommand{\ddb}{\ddot{b}}
\newcommand{\ddn}{\ddot{n}}
\newcommand{\ba}{\begin{array}}
\newcommand{\ea}{\end{array}}
\def\be{\begin{equation}}
\def\ee{\end{equation}}
\def\bea{\begin{eqnarray}}
\def\eea{\end{eqnarray}}
\def\bs{\begin{subequations}}
\def\es{\end{subequations}}
\def\g{\gamma}
\def\G{\Gamma}
\def\vp{\varphi}
\def\mpl{M_{\rm P}}
\def\ms{M_{\rm s}}
\def\ls{\ell_{\rm s}}
\def\lp{\ell_{\rm pl}}
\def\l{\lambda}
\def\gs{g_{\rm s}}
\def\d{\partial}
\def\co{{\cal O}}
\def\sp{\;\;\;,\;\;\;}
\def\spa{\;\;\;}
\def\r{\rho}
\def\dr{\dot r}
\def\dt{\dot\varphi}
\def\e{\epsilon}
\def\k{\kappa}
\def\m{\mu}
\def\n{\nu}
\def\om{\omega}
\def\tn{\tilde \nu}
\def\p{\phi}
\def\vp{\varphi}
\def\P{\Phi}
\def\r{\rho}
\def\s{\sigma}
\def\t{\tau}
\def\x{\chi}
\def\z{\zeta}
\def\a{\alpha}
\def\b{\beta}
\def\de{\delta}
\def\bra#1{\left\langle #1\right|}
\def\ket#1{\left| #1\right\rangle}
\newcommand{\stt}{\small\tt}
\renewcommand{\theequation}{\arabic{section}.\arabic{equation}}
\newcommand{\eq}[1]{equation~(\ref{#1})}
\newcommand{\eqs}[2]{equations~(\ref{#1}) and~(\ref{#2})}
\newcommand{\eqto}[2]{equations~(\ref{#1}) to~(\ref{#2})}
\newcommand{\fig}[1]{Fig.~(\ref{#1})}
\newcommand{\figs}[2]{Figs.~(\ref{#1}) and~(\ref{#2})}
\newcommand{\GeV}{\mbox{GeV}}
\def\ricci{R_{\m\n} R^{\m\n}}
\def\riemann{R_{\m\n\l\s} R^{\m\n\l\s}}
\def\triemann{\tilde R_{\m\n\l\s} \tilde R^{\m\n\l\s}}
\def\tricci{\tilde R_{\m\n} \tilde R^{\m\n}}
\title{Scale dependence of the UHECR neutrino flux in extra-dimension models}
\author{K.R.S. Balaji$^1$ \email[Email:]{balaji@hep.physics.mcgill.ca}
and Jukka Maalampi$^2$ \email[Email:]{maalampi@cc.jyu.fi}}
\affiliation{ $^1$ Department of Physics, McGill University, Montr\'eal, QC, 
Canada H3A 2T8}
\affiliation { $^2$ Department of Physics, Jyv\"askyla University, Jyv\"askyl\"a,
Finland}
\affiliation { Helsinki Institute of Physics, Helsinki, Finland}
\begin{abstract}
Ultra high energy cosmic ray (UHECR) neutrino fluxes measured in a fixed target 
detector can have a scale dependence. In the usual standard model or any 
extensions of this model (which are renormalizable),  the effect is 
observationally very small.  However, this need not be the case in models with  
extra-spatial dimensions, where the neutrino mass parameter can receive large 
corrections due to a power-law running. Hence, the scale dependence may lead to a 
measurable deviation from  the standard prediction for the neutrino flux ratio. 
\end{abstract}
\pacs{98.80.Cq.}
\maketitle
\section{Introduction}
In the last two decades, one of the significant success of particle physics has 
been 
the 
confirmation of a neutrino anomaly, both in the solar and atmospheric sector 
\cite{sk}. 
The most likely solution to the anomaly is to introduce a small neutrino
mass and hence the notion of neutrino oscillation \cite{ponty} which is similar in 
spirit to
quark sector. Recent atmospheric neutrino data has indeed shown an observable dip 
in its 
zenith 
angle spectra as expected for massive neutrino oscillation \cite{atm}. In a 
realistic three 
flavor analysis,  an important part of the solution amounts to finding the allowed 
parameter space for the
the mixing angles; the solar $(\theta_S)$, the atmospheric $(\theta_A)$ and a
reactor angle $(\theta_R)$. The best fit values for these three mixing parameters 
seem to
indicate a pattern spanning from almost being negligible to moderate to maximal 
\cite{nuph}. 
In the case of solar neutrinos, the mixing of $\nu_e$ with active neutrinos has 
the central
value such that $\tan \theta_S \approx 0.7$ (moderate). In contrast, the 
atmospheric
mixings involving largely of $\nu_\mu$ prefers $\tan\theta_A \approx 1$ (maximal) 
while the
reactor angle, which determines the relative proportion of $\nu_e$ in the heaviest 
mass 
eigenstate, is consistent with zero mixings, $\tan\theta_R \ll 1$ \cite{chooz}. 

The fact that $\tan\theta_A \approx 1$ has an important consequence for
the neutrino fluxes which are ultra-relativistic in energies. It has long been 
realized
that UHECR neutrinos (which are expected to be sourced by cosmic objects such as 
AGNs) when
measured by ground based detectors, the expected flavor 
ratio $\phi_e:\phi_\mu:\phi_\tau=1:1:1$ 
\cite{jukka}. Henceforth, we shall call this expectation as the {\em bench mark} 
value. 
This prediction is important for at least three fundamental reasons: (i) it forms 
an 
independent verification of the neutrino parameters which are phenomenologically 
extracted
from solar, atmospheric and reactor data, (ii) it is has been realized to be a 
test bed for 
some interesting new physics predictions (decay, pseudo-Dirac splittings, 
active-sterile
mixings) which are not yet
resolved \cite{jb} and (iii) it could provide further opportunity in our 
understanding of 
fermion mixings and masses; for instance, are there any fundamental symmetries in 
the
$\mu-\tau$ block which leads to maximal mixings. It is expected that several of 
the upcoming neutrino telescopes \cite{nutel} will be tuned to verify the {\em 
bench mark}
value besides looking for many of the new signatures mentioned here. 

 In the present analysis, we point out that the scale dependence of the neutrino 
parameters
 can also be a source which alters the {\em bench mark} expectations. In 
scattering processes
 involving UHECR neutrinos, the momentum transfer square $\mu$ is expected to 
saturate at $10^4$ GeV$^2$ beyond which point there is a strong energy suppression 
\cite{reno}. It is well known that at this scale, the effects of running on 
neutrino mixings are very small \cite{rge}. However, this need not be the case in 
models with extra-spatial dimensions, thereby, leading to modifications to the 
 {\em bench mark} values. This forms the main theme of our analysis. 
\section{standard lore}
   It is instructive to first review the standard {\em bench mark} expectations. 
Massive
 neutrinos,
similar to quarks, have two eigenbasis, the flavor $(\nu_\alpha)$ and mass 
eigenbasis 
$(\nu_i)$ with corresponding mass eigenvalues, $m_i$.
A unitary matrix relates the two basis, such that $\nu_\alpha = U_{\alpha i} 
\nu_i$ 
where, the summation over the mass eigenstates is assumed. In the limit of small
mixings, one could define the angles in the following manner. $\theta_S$ mixes
states $\nu_1$ and $\nu_2$, $\theta_A$ mixes $\nu_2$ and $\nu_3$ and $\theta_R$ 
mixes $\nu_1$ and $\nu_3$. In this notation, without loss of generality, we can 
assume 
a hierarchy of states, where, $m_1 < m_2 < m_3$ and the relevant solar and 
atmospheric 
splittings, 
are $\Delta_S = m_2^2 -m_1^2$ and $\Delta_A = m_3^2 - m_2^2$ respectively. In the 
case of UHECR neutrinos which travel astronomical distances,  the coherence 
between the various mass eigenstates is averaged out. As a result, once these 
neutrinos 
are produced, they essentially travel (galactic distances) as individual mass 
eigenstate, until at the point of detection. In a ground based detector, the 
probability of 
measuring a UHECR neutrino of a given flavor is then given as
\bea
\phi_e &=& 1 + 2x(2c_A^2-1)~;~x=(s_Sc_S)^2~,\nonumber\\
\phi_\mu &=& 2xc_A^2 + 2(c_A^4(1-2x) +s_A^4)~,\nonumber\\
\phi_\tau &=& s_{2A}^2 + 2xs_A^2(1-c_A^2)~,
\label{flux}
\eea
where $s$ and $c$ denote sine and cosine, respectively. It is clear from the above 
expression, 
that maximal atmospheric mixing leads to the conclusion that all neutrino flavors
must be detected with the same weight factor. In deriving this result, we have 
disregarded
the mixing corresponding to reactor experiments, which is consistent with zero 
\cite{chooz}. 
Given this result, we shall consider the modifications that may alter the 
prediction in
(\ref{flux}) for $\theta_A \neq \pi/4$.
\section{ renormalization group effects}
UHECR neutrinos incident on a target material can undergo both charged and neutral 
current scattering processes. The individual neutrino flavor states $\nu_\alpha$  
are derived by folding the matrix element $U_{\alpha i}$ corresponding to the 
incident mass eigenstate $\nu_i$. In a scattering process, which involves large 
momentum transfers, the mixing matrix element $U_{\alpha i}$ can pick up a scale 
dependence. However, in practice, the momentum transfer square $(\mu)$ saturates  
at $\mu \sim 10^4$ GeV$^2$ beyond which the cross section is damped \cite{reno}. It 
is well known that for scales around 
this value, the effects of neutrino mass running is negligible \cite{rge}. As 
mentioned earlier, this need not be the case if we consider models with 
extra-space dimensions. Furthermore, in this case, depending on the the mass of KK 
excitation for the gauge boson $\mu$ can saturate at a much higher value. Present 
collider bounds suggest that the lowest KK excited state can have a mass $\sim few 
~100$ GeV \cite{gl} leading to $\mu \sim 1$ TeV as the scale of extra-dimension.

In the following, we consider a class of models where the neutrinos are localized 
in the brane, such that for $\delta$ extra spatial dimensions and for scales 
$\Lambda > \tilde \Lambda$  (electroweak scale) we have the evolution equation  
for the mass parameter \cite{dudas}
\bea
16 \pi^2 \frac{d\kappa}{d \ln \Lambda} &=& (-3 g_2^2 + 2 \lambda + 2S) t_\delta
\nonumber\\ &-&\frac{3t_\delta \kappa}{2}[(Y_l^\dagger Y_l) + (Y_l^\dagger 
Y_l)^T]~,
\nonumber\\
S &=& \mbox{Tr}(3Y_u^\dagger Y_u +3Y_d^\dagger Y_d +Y_l^\dagger Y_l)~,\nonumber\\
t_\delta &=& (\frac{\Lambda}{\tilde\Lambda})^{\delta}X_\delta~;~X_\delta = 
\frac{2}{\delta}
\pi^{\delta/2}\Gamma(\delta/2)  ~.
\label{edrun1}
\eea
 In (\ref{edrun1}) $Y_{u,d,l}$ are the up-quark, down-quark 
and charged lepton Yukawa couplings. For our purposes, we will focus on the 
contributions due to the charged lepton Yukawa couplings such that integrating 
(\ref{edrun1}) yields
\bea
\ln (\frac{\kappa}{\tilde\kappa}) 
&=& \frac{3Y_l^2}{16\pi^2}(1-(\frac{\Lambda}{\tilde\Lambda})^\delta) 
\frac{X_\delta}{\delta}
 \equiv \eta_l~.
\label{edrun2}
\eea
\begin{figure}
\centerline{\epsfxsize=3.0 in \epsfbox{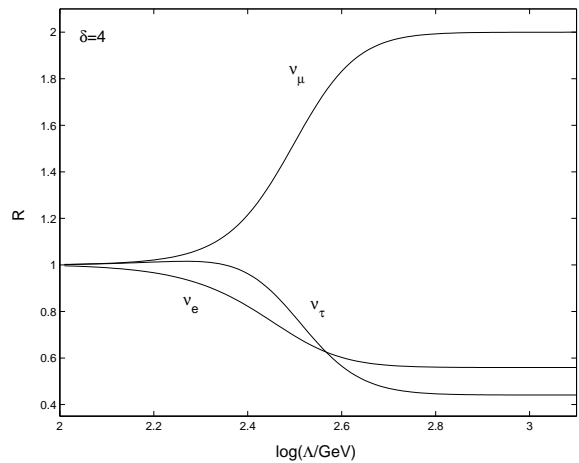}}
\caption{Variation of the flux ratio $R$ with scale for $\delta = 4$.}
\label{fig1}
\end{figure}
 It is important to note that in (\ref{edrun2}) the 
nature of  running depends strongly on the value for $\delta$. As a result, the 
mixings can run significantly even for a small variation in the scales. This 
arises from the power law running which can compensate for the energy suppression 
in the propagator for an off-shell neutrino.   Alternatively, the energy 
enhancement is due to the  multiplicity factor (which is 
$(\Lambda/\tilde\Lambda)^\delta$ ) and arises from the number of  Kaluza-Klein 
states, which for a given $\Lambda$ and $\delta$ can be large.
 
 The running of the masses translates to a running of the neutrino mixings. We 
estimate the corrections to the leading order in the enhancement (essentially
$t_\delta$) although this expansion need not be perturbative, especially for large 
$\mu$ and/or
$\delta$. Also, we do not write down the corrections other than due to $Y_l$ since 
it is not relevant to our discussion. Following (\ref{edrun2})  up to  $O(\eta_l)$  
the change in mass matrix element as a function of scale is obtained to be

\be
\kappa_{\alpha\beta}(\Lambda) \simeq \tilde\kappa_{\alpha \beta} (\tilde\Lambda)(1 
+ \eta_l) 
\equiv \tilde \kappa_{\alpha\beta} (1+\eta_l)~.
\label{edmatele}
\ee 

In (\ref{edmatele}) $\tilde\kappa_{\alpha \beta}(\tilde\Lambda)$ is taken to be 
the value of the element at the electroweak scale. In the limit of two flavor 
mixing (which can be arranged
if $U_{e3}=0$)  following (\ref{edmatele}) the mixing angle depends on scale in 
the form

\be
\tan\theta_{\alpha \beta}(\Lambda) = 
\frac{2\kappa_{\alpha \beta}(\Lambda)}{\kappa_{\alpha \alpha}(\Lambda) - 
\kappa_{\beta \beta}(\Lambda)}~ \approx 
\tan\tilde\theta_{\alpha\beta}(\tilde\Lambda) 
(1- \eta_l)~.
\label{edmixang}
\ee
 Having obtained this change in the mixing angle (up to first order in $\eta_l$) 
we can consider
the change in $\theta_A$ for which case, we identify $\alpha = \mu$ and $\beta = 
\tau$ and
take $Y_l = Y_\tau$. We plot the modification to the flavor fluxes as shown in 
Fig.1 where we choose $\delta =4$.  To be specific, we have assumed MSSM Yukawa 
couplings for the tau lepton at $\tan\beta=50$. In the plot, we show the variation 
for flux $R$ with scale and as $\Lambda$ increases, $\theta_A \to 0$. As we should 
expect, in this limit, the muon flux approaches a value which is consistent with 
no $\nu_\mu \leftrightarrow \nu_\tau$ mixing. Infact, at this energy scale, an 
observation  (if done) of the muon flux will
constitute a direct measurement of the flux at the point of production (modulo the 
small errors due
to $U_{e3}\neq 0$.) In this simple exercise, our choice for $\delta$ is purely for 
illustrative purposes since, it is a free parameter and can be fixed depending on 
the cross section strength required for an observable effect.
 \subsection{An example of a $2\to 3$ scattering process}
  We now consider a physical process where it might be possible to have a 
measurement of the scale dependence along with a unique signature. Essentially, we 
are considering a $2\to 3$ tree level scattering process whose  Feynman graph is 
shown in Fig.2. In this process, a deeply virtual neutrino $(\nu^*)$ eventually 
fragments to a gauge boson $(G)$ and an accompanying lepton: $\nu^* \to G + 
~\mbox{leptons}$. 
The final state gauge bosons can be identified via their decay jets. This process 
is very similar to the electroproduction of heavy Majorana neutrinos considered 
earlier by 
Buchm\"uller and Greub \cite{buch}. We remind that in our case, the state $G$ can 
also include KK excitations, hence, unlike in the standard model case, $\mu $ can 
saturate at values larger than $10^4$ GeV$^2$. 
\begin{figure}
\centerline{\epsfxsize=3.0 in \epsfbox{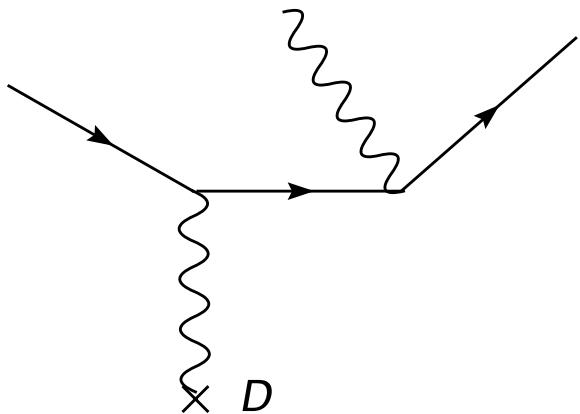}}
\caption{Feynman graph showing an incident neutrino scattering off the detector 
$D$  followed by a  virtual neutrino state fragmenting in to a gauge boson (wiggly 
line) and a final state lepton.}
\label{fig2}
\end{figure}

In the following, we outline the feasibility of measuring the process, while, a 
detailed calculation is beyond the scope of this paper. In principle, we wish to 
show that the process may not encounter the usual propagator suppression for $\mu 
\gg 1$. Let us examine the off-shell neutrino propagator in this energy regime. 
The relevant part of interest in the  propagator reads as
 \bea
 D(\mu) = \frac{\kappa_{\alpha\beta}}{\mu-\kappa_{\alpha\beta}^2 } &\simeq 
&\tilde\kappa_{\alpha\beta}
 (\frac{\Lambda}{\tilde\Lambda }) ^\delta  \frac{1}{\mu - 
\kappa_{\alpha\beta}^2}+...~,\nonumber\\
 &\simeq& \tilde\kappa_{\alpha\beta}
 (\frac{\Lambda}{\tilde\Lambda })^{\delta}\frac{\tilde\Lambda^{\delta}}{\mu 
\tilde\Lambda ^{\delta}- \Lambda^{\delta} \tilde\kappa_{\alpha\beta}^2 } +...~.
 \label{edprop}
 \eea
where $...$ denote higher order corrections to $\kappa_{\alpha\beta}$.  We 
consider the possibility where the scale of extra-dimensions is within the range 
of experimental reach such that for some allowed $\mu$ we have  $\Lambda \sim 
\sqrt{|\mu|}$. In this case, depending on the value of $\delta$ we should expect 
the cross section to  grow with energy. Clearly, from (\ref{edprop})  we find 
\be
 D(\mu \gg 1)  \sim \frac{\tilde\kappa_{\alpha\beta}}{\tilde\Lambda ^\delta} 
 |\mu|^{\delta/2 -1}~.
  \label{edprop1}
 \ee
Note that from (\ref{edprop1}) for $\delta =0$ we reproduce the expected energy 
suppression as in conventional non-extra-dimensional models. Thus, in all such 
theories, neutrinos which are emitted off the gauge boson vertex will always 
prefer to be on-shell.  For $\delta \neq 0$ we find that the theory shows the 
usual pathology of cross section growing with energy \cite{maltoni}. This becomes 
severe as $\delta$ increases. Therefore, as $\delta$  increases, even for scales 
not too far from $\tilde\Lambda$ the cross section can grow significantly with 
energy. This also reflects the fact that the theory is unitarity violating. 
However, we also need to ensure that there are no low-energy anomalous processes 
which might be in conflict with the standard model results \cite{sandip}. For 
instance, neutrino-nucleon cross sections which violate unitarity can have
observable anomalous cross sections in the corresponding low-energy elastic 
processes \cite{goldberg}. Alternatively, one can examine the effects of new 
physics on final state 
interactions for a given process. If new physics occurs at the TeV scale, then an 
observable deviation of $\sim (0.01\%)$ is expected for scattering processes at 
the 
electroweak scale \cite{domo}. Currently, this small deviation is consistent with 
the
LEP limits. However, we note that isolating any anomalous events 
may be an experimental challenge, especially, due to a lack of knowledge on the
parton distribution functions involving states in the continuum.

 In conclusion, the present analysis does demonstrate a possible window to observe 
the scale dependence of UHECR neutrino fluxes. We have taken a representative set 
of low energy neutrino parameters and analyzed the evolution of the mixing with 
scale. It might be of
interest perform a more general analysis where we also consider the running of the 
CP phases 
and the solar mixing as well. An important ingredient in estimating the running is 
the value of neutrino parameters at the electroweak scale. Fortunately, we already 
have a good idea about the neutrino parameters $(\theta_{S,A},~\Delta_{S,A})$ from 
some very accurate phenomenological
analysis
of the solar and atmospheric data \cite{nuph}. Contrary to non-extra dimensional 
models where neutrino mass degeneracy is an important ingredient; 
extra-dimensional models
may relax this requirement since power-law running can account for large radiative 
corrections. 
As we have shown, the variation to the {\em bench mark} values could already occur 
for scales
not too far from the electroweak scale.
This implies that if the scale of extra-dimensions is within the reach of the 
neutrino telescopes, (then independent of the nature of the neutrino spectra), the 
effect which we predict should be observed.  In addition, a measurement of scale 
dependence can also carry some unique and interesting signals, like the one 
described in the $2 \to 3$ scattering process.

{\bf Acknowledgements}:
The work of KB is funded by NSERC (Canada) and by the Fonds de Recherche sur la 
Nature 
et les Technologies du Qu\'ebec. He also thanks the support and hospitality at JU 
( Finland)
 where this work was completed. JM is supported by funds from the Academy of 
Finland under contract 104915 and 107293. We thank Rabi Mohapatra, Harry Lam, Guy 
D. Moore, Sandip Pakvasa and Kimmo Kainulainen for useful discussions.
 
\end{document}